\documentclass[12pt]{article}

\usepackage{geometry}
\usepackage{graphicx}
\usepackage[usenames,dvipsnames]{color}
\usepackage{color}
\usepackage[english]{babel}
\usepackage{epstopdf}
\usepackage{latexsym}
\usepackage{hyperref}
\usepackage{url}
\usepackage{amstext}
\usepackage{amssymb}
\usepackage{amsmath}
\usepackage{url}

\hypersetup{
	colorlinks=true,
	linkcolor=blue,
	filecolor=magenta,
	urlcolor=blue,
}
\usepackage{times}
\usepackage{hyperref}
\usepackage{url}
\usepackage{wrapfig}
\usepackage{float}
\usepackage{cite}
\usepackage{tikz}
\usepackage{pgfplots}
\usepackage{caption}
\usepackage{subcaption}
\usepackage{lineno}
\newcommand*\patchAmsMathEnvironmentForLineno[1]{%
  \expandafter\let\csname old#1\expandafter\endcsname\csname #1\endcsname
  \expandafter\let\csname oldend#1\expandafter\endcsname\csname end#1\endcsname
  \renewenvironment{#1}%
     {\linenomath\csname old#1\endcsname}%
     {\csname oldend#1\endcsname\endlinenomath}}%
\newcommand*\patchBothAmsMathEnvironmentsForLineno[1]{%
  \patchAmsMathEnvironmentForLineno{#1}%
  \patchAmsMathEnvironmentForLineno{#1*}}%
\AtBeginDocument{%
\patchBothAmsMathEnvironmentsForLineno{equation}%
\patchBothAmsMathEnvironmentsForLineno{align}%
\patchBothAmsMathEnvironmentsForLineno{flalign}%
\patchBothAmsMathEnvironmentsForLineno{alignat}%
\patchBothAmsMathEnvironmentsForLineno{gather}%
\patchBothAmsMathEnvironmentsForLineno{multline}%
\patchBothAmsMathEnvironmentsForLineno{multline}%
}
\DeclareRobustCommand{\hwplotAA}{\raisebox{2pt}{\tikz{\draw[blue,solid,line width=1.5pt](0,0)--(4mm,0);}}}

\DeclareRobustCommand{\hwplotBA}{\raisebox{2pt}{\tikz{\draw[blue,solid,line width=1.5pt](0,0)--(4mm,0);}}}
\DeclareRobustCommand{\hwplotBB}{\raisebox{2pt}{\tikz{\draw[magenta,solid,line width=1.5pt](0,0)--(0.75mm,0);}}}

\title{Effect of light intensity on resonance patterns in CDIMA reaction}
\date{}
\author{Swadesh Pal$^{1}$ and Malay Banerjee$^{1}$\thanks{Corresponding author: E-mail: malayb@iitk.ac.in, Ph. +91-512-679-6157, Fax. +91-512-679-7500}}

\begin{document}

\maketitle
\vspace*{-1cm}
\begin{center}
$^{1}$Department of Mathematics and Statistics,
IIT Kanpur, Kanpur, India, 208016
\end{center}

\begin{abstract}
Photosensitive CDIMA reaction-diffusion equation is considered to explain the resonance in the linearly coupled system. The conditions for Turing instability is obtained for the coupled reaction-diffusion system. Also, determining the critical diffusion coefficients for exiting the resonance type pattern is also discussed with analytical conditions. The photosensitive effect on $3:1$ resonance forcing is studied in the presence of strong and weak couplings. Numerical simulations are performed to validate the theoretical findings. 

\end{abstract}

\section{Introduction}

Spatio-temporal patterns are widely observed in the chemical \cite{One,Two}, biological \cite{Fourteen}, and physical systems \cite{Sixteen}. Turing \cite{Seventeen} first proposed the reaction-diffusion mechanism which explains that a homogeneous stable steady-state can be destabilized through heterogeneous pattern and is capable to develop spatially heterogeneous stationary pattern. The instability of the homogeneous steady-state triggered by random disturbances around homogeneous steady-state. In theoretical approach, variety of spatio-temporal models are proposed which are capable to produce Turing patterns with their own characteristics and different reaction kinetics.

The experimental evidence for Turing pattern was first obtained by Castests \textit{et al.} \cite{Four}, in chloride-iodide-malic acid (CIMA) reaction, and later on in the chlorine dioxide-iodine-malonic acid (CDIMA) reaction \cite{Five,Six}.  The Turing pattern in chemical system is also obtained via experimental with CIMA reaction within spatial reactor (see P. De Kepper \textit{et al.}\cite{Nine}, Q. Ouyang \textit{et al.} \cite{Twelve}, and B. Rudovics \textit{et al.}\cite{Thirteen}). Further, A. K. Horvath \textit{et al.} \cite{Six} shown that the Turing structures can be controlled through visible light and examined the roles of frequency and intensity of periodic illumination in suppressing Turing patterns .

The dynamics of such type model with CDIMA reaction kinetics exhibits several types of stationary patterns: labyrinthine, hexagons, and black eye \cite{Twelve}. A. P. Munuzuri \textit{et al.} \cite{One} shown that CDIMA reaction is photosensitive to visible light. They proved experimentally as well as computationally that in the presence of visible light CDIMA reaction suppresses oscillations for the starch free system but strongly shifts to the steady-state concentrations in a system with starch. Previous results by D. G. Miguez \textit{et al.} \cite{Eleven} have shown that the experimental and numerical results of two "two-component" interacting system is capable to produce Turing patterns in coupled layers with diffusion, with limitation on the light intensity in the second coupled system. Based on Galerkin method, P. Ghosh \textit{et al.} \cite{Ten} theoretically analysed the light-induced Turing patterns of the CDIMA reaction, and the transition behaviour by varying the illumination intensity.

Studies on coupling and synchronization is considered to be one of the developing area in nonlinear dynamics of spatial pattern formation strongly motivated by both biological and theoretical aspects. In \cite{Eleven}, authors have considered the effect of light intensity only on the first system rather than the second one and also they have used different vertical diffusion terms. Also, in \cite{Eighteen}, authors have considered the light intensity with respect to both the space variables. Until recently, authors have concentrated on determining the Turing instability condition by choosing diffusion coefficients for the coupled systems as bifurcation parameter. In general, for the models without coupling, authors have fixed one of the diffusion coefficient to a constant value and then varied other diffusion coefficient for which the coupled system becomes unstable.

Reaction-diffusion systems composed of spatially interacting layers have received much attention over the past decades. Also most experimental studies of Turing patterns have explored structures with a single characteristic wavelength. An interesting exception is that the black eye pattern, interpreted as a resonance between two hexagonal lattices \cite{Three}. Resonant patterns \cite{Fifteen} has been observed in reaction-diffusion equation with forcing terms or linearly coupled systems. In \cite{Eight}, authors have explored numerically various types of resonance forcing e.g. $\sqrt{3}:1,2:1,3:1$, and so on.

In this paper, we present the theoretical results to determine critical diffusion coefficients and perform numerical simulations for a two coupled layers interacting Turing system via diffusion parameters. Taking the advantage of photosensitive characteristics of the CDIMA reaction, we study the interacting Turing pattern with different Turing modes. Also, we report the effect of the illumination on the resulting pattern due to the strong coupling and weak coupling. The numerical results show how light intensity effects to the resonance pattern with $3:1$ forcing.

\section{The Model and Method}

Consider the Lengyel-Epstein two-variable reaction-diffusion model \cite{Two}, which includes the effect of illumination \cite{One}
\begin{equation}{\label{T1}}
\begin{aligned}
\frac{\partial u}{\partial t} & = d_{1}\Delta u+a-u-\frac{4uv}{u^{2}+1}-w, \\
\frac{\partial v}{\partial t} & = \sigma\bigg{[}d_{2}\Delta v+b\bigg{(}u-\frac{uv}{u^{2}+1}+w\bigg{)}\bigg{]},
\end{aligned}
\end{equation}
with no-flux boundary conditions and feasible initial conditions. We take small perturbation around the homogeneous steady-state as initial condition. Here $u$ and $v$ are the dimensionless concentrations of $I^{-}$ and $ClO_{2}^{-}$, respectively; $a,b$ and $\sigma$ are dimensionless parameters and $w$ is also dimensionless parameter, which is proportional to the light intensity.

We suppose that \begin{align*}
f(u,v) &= a-u-\frac{4uv}{u^{2}+1}-w,\\
g(u,v) &= \sigma b\bigg{(}u-\frac{uv}{u^{2}+1}+w\bigg{)}.
\end{align*}

Now, we derive the Turing instability conditions for the system (\ref{T1}). Turing instability condition of a reaction-diffusion system is based on the homogeneous steady-state of the system and it says that the homogeneous steady-state is stable without diffusion and is unstable due to small amplitude heterogeneous perturbation for the system with diffusion. The homogeneous steady-state of the system (\ref{T1}) are given by the solutions of the equations $f(u,v)=0$ $\&$ $g(u,v)=0$ $\Rightarrow u=\frac{a}{5}-w, v=\frac{a}{5}\frac{u^{2}+1}{u}$. For feasibility of the steady-state we need $a > 5w$. $(u^{*},v^{*})$ is the equilibrium point of the temporal model corresponding to the spatio-temporal (\ref{T1}). Sign of the Jacobian matrix at $(u^{*},v^{*})$ is
 $\begin{bmatrix}
+ & - \\
+ & -
\end{bmatrix}.$ Also, the determinant of the Jacobian matrix at the equilibrium point is always positive. So, the equilibrium point is stable if
$$b>\frac{3a^{3}-25aw(a-w)+125(w^{3}-a+w)}{5\sigma (a-5w)^{2}}.$$

We assume that $(u^{*},v^{*})$ is stable. Now, $u=u^{*}$ and $v=v^{*}$ is a homogeneous solution of the system (\ref{T1}). Based on the diffusion coefficients the homogeneous solution of the system (\ref{T1}) may be stable or unstable. We need to find out the conditions in terms of $d_{1}$ and $d_{2}$ such that the homogeneous steady-state becomes unstable. In order to determine the Turing instability condition, the system (\ref{T1}) is linearised around $(u^{*},v^{*})$ and we get
\begin{equation}{\label{T2}}
\frac{\partial \zeta}{\partial t}  = \mathbf{D}\Delta \zeta +\mathbf{A} \zeta,
\end{equation}
where $\zeta = (u-u^{*},v-v^{*})^{T}$ is small heterogeneous perturbation around homogeneous steady-state, $\mathbf{D}=\mbox{diag}\{ d_{1}, \sigma d_{2}\}$ and $\mathbf{A}=
\begin{bmatrix}
f_{u} &     f_{v} \\
g_{u}    & g_{v}
\end{bmatrix}$ evaluated at $u=u^{*}$ and $v=v^{*}.$

Let $\zeta=e^{\lambda t+i(lx+my)}\mathbf{\Psi}$ be the solution of (\ref{T2}) and then substituting in (\ref{T2}) and we get:
\begin{equation}{\label{T3}}
\lambda \mathbf{\Psi} =(\mathbf{A}-k^{2}\mathbf{D})\mathbf{\Psi},~\mbox{where}~k^{2}=l^{2}+m^{2}.
\end{equation}

Here we are interested to find the non-trivial solutions of (\ref{T3}), so $\lambda$ is a solution of the characteristic equation
\begin{equation}{\label{T4}}
\mathfrak{D} (k^{2},\lambda)=\mbox{det}(\mathbf{A}-k^{2}\mathbf{D}-\lambda \mathbf{I}_{2})=0.
\end{equation}

\begin{figure}
    \begin{center}
    \includegraphics[width=0.45\textwidth]{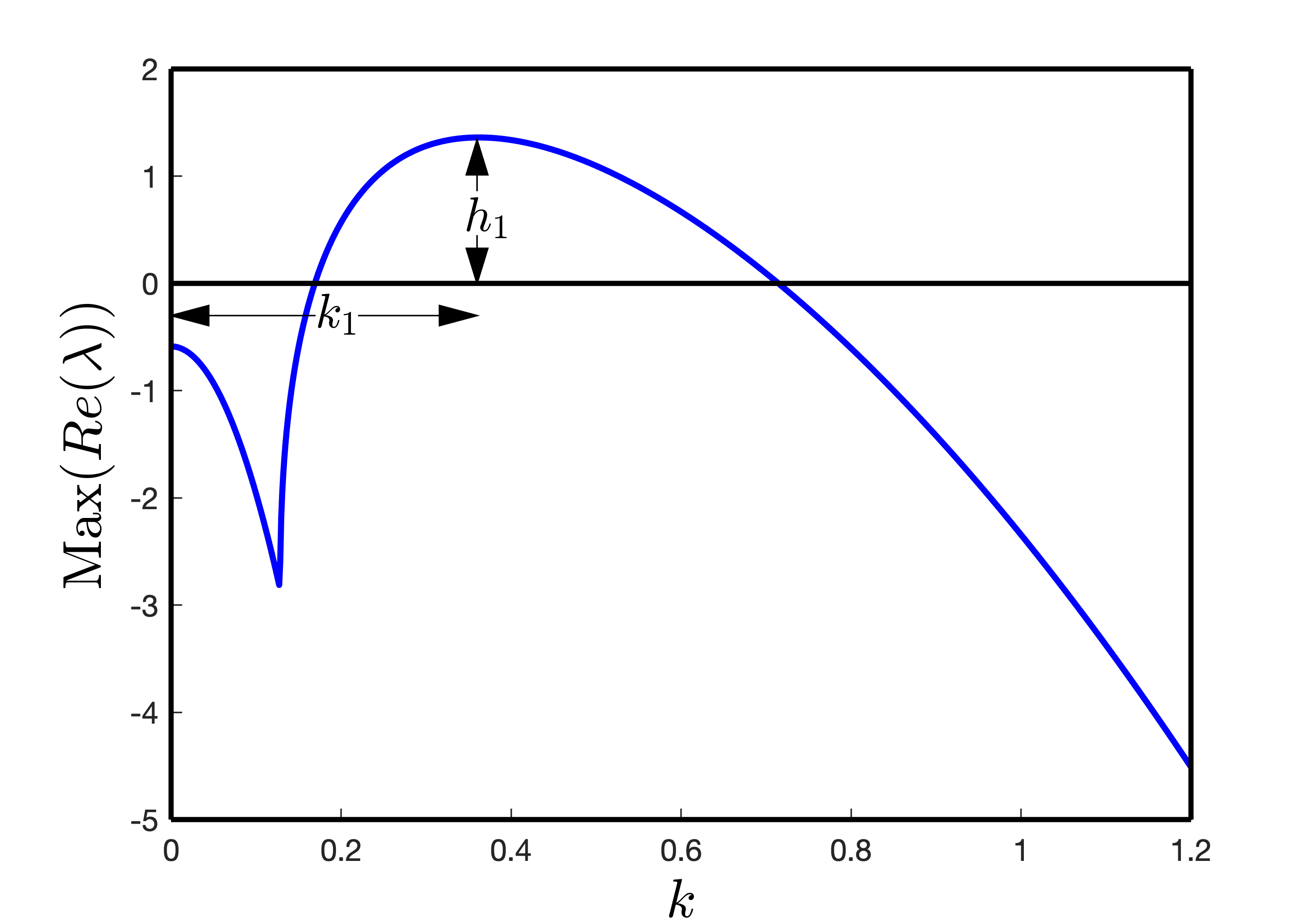}
    \end{center}
     \caption{Blue curve (\hwplotAA) represents the maximum of $Re(\lambda)$ with respect to $k$.}
     \label{fig:TF1}
\end{figure}

The stability and instability of $(u^{*},v^{*})$ are related to the sign of $Re(\lambda)$ and solely related to the parameter $d_{1}$ and $d_{2}$. For $k=0$, we have the real part of two eigenvalues are negative. If $\mbox{max}(Re(\lambda))<0, \forall k$, then the homogeneous steady-state for system (\ref{T1}) is stable. So, for instability we need to choose $d_{1}$ and $d_{2}$ in such a way that $Re(\lambda(k))>0$ for a range of values of $k$. Suppose at $k=k_{1}$, $\lambda$ has a real maximum and the maximum value is $h_{1}$ [see Fig. \ref{fig:TF1}]. Therefore, $\mathfrak{D} (k_{1}^{2},h_{1})=0$ and $\lambda'(k_{1})=0$. In order to find related values of $d_{1}$ and $d_{2}$ we can assume the desired values of $h_{1}$ and $k_{1}$ and then solve two equations to find $d_{1}$ and $d_{2}$. After solving these two equations, we can derive the values of $d_{1}$ and $d_{2}$. It is important that the desired valued of $h_{1}$ and $k_{1}$ should be reasonable and corresponding values of $d_{1}$ and $d_{2}$ are feasible.  So, for this $d_{1}$ and $d_{2}$, we can write $\mbox{max}(Re(\lambda))=h_{1}$ at $k=k_{1}$ and the pair $(k_{1},h_{1})$ is called `Turing mode'.

In order to study the internal spatial resonance pattern, we can construct a suitable model with two interacting Turing modes by coupling two systems of the form (\ref{T2}) linearly, each having a single Turing mode \cite{Eight}. Physically, such a model represent two thin layers of gel that meet at an interface. Each layer contains the same set of reactants with the same kinetics; the difference between the layers arise from the difference between the diffusions due to physical of chemical factors. The interaction between two such coupled layers can be described by a set of four partial differential equations:
\begin{equation}{\label{1}}
\begin{aligned}
\frac{\partial u_{r}}{\partial t} & = D_{u_{r}}\Delta u_{r}+a-u_{r}-\frac{4u_{r}v_{r}}{u_{r}^{2}+1}-w + \alpha (u_{s}-u_{r}), \\
\frac{\partial v_{r}}{\partial t} & = \sigma\bigg{[}D_{v_{r}}\Delta v_{r} +b\bigg{(}u_{r}-\frac{u_{r}v_{r}}{u_{r}^{2}+1}+w\bigg{)}\bigg{]}+ \beta (v_{s}-v_{r}),
\end{aligned}
\end{equation}
where $u_{r}$ and $v_{r}$ are the dimensionless concentrations of $I^{-}$ and $ClO_{2}^{-}$ and their diffusion coefficients are $D_{u_{r}}$ and $D_{v_{r}}$ respectively and distinguished by the subscripts $r,s=1,2,r\neq s$ with no flux boundary condition and the initial condition as described earlier. Here $\alpha$ and $\beta$ represents the linear coupling strength of two layers, $f$ and $g$ are the same reaction kinetics as defined above.

\begin{figure}[H]
\begin{center}
        \centering
        \includegraphics[width=0.6\textwidth]{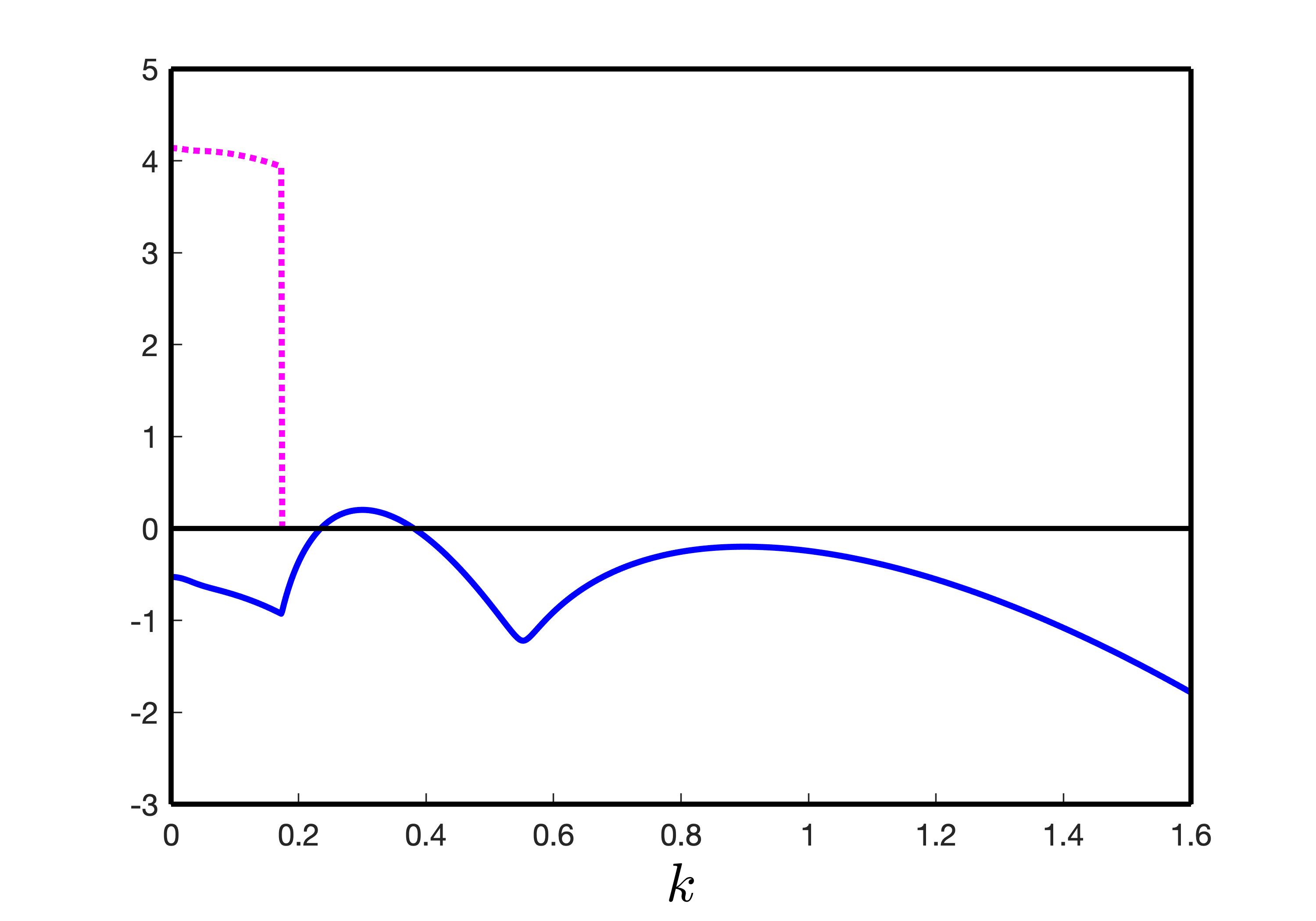}
        \caption{Solid curve (\hwplotBA) represents the $\mbox{max}(Re(\lambda))$ and the dotted curve (\hwplotBB \hwplotBB \hwplotBB) is the imaginary part of $\lambda$ corresponding to $\mbox{max}(Re(\lambda))$.}\label{fig:TF2}
\end{center}
\end{figure}

Now we derive the Turing instability condition for the system (\ref{1}). The homogeneous steady-state of the system (\ref{1}) is $u_{2}^{*}=u_{1}^{*}=u^{*}$ and $v_{2}^{*}=v_{1}^{*}=v^{*}$. Suppose, $U^{*}=(u_{1}^{*},v_{1}^{*},u_{2}^{*},v_{2}^{*})$ is a steady-state solution of (\ref{1}) and also $U^{*}$ is stable equilibrium of the temporal model corresponding to the spatio-temporal system (\ref{1}).  Linearising (\ref{1}) around $U^{*}$, we obtain the system of linear equations in terms of the perturbation variable $\xi$,
\begin{equation}{\label{4}}
\frac{\partial \mathbf{\xi}}{\partial t}  = \mathbf{D}\Delta \mathbf{\xi} +\mathbf{F} \mathbf{\xi},
\end{equation}
where $\mathbf{\xi} = (u_{1}-u_{1}^{*},v_{1}-v_{1}^{*},u_{2}-u_{2}^{*},v_{2}-v_{2}^{*})^{T}$, $\mathbf{D}=\mbox{diag}\{ D_{u_{1}}, \sigma D_{v_{1}}, D_{u_{2}}, \sigma D_{v_{2}}\}$, and
\begin{align*}
\mathbf{F}=\begin{bmatrix}
f_{u_{1}}-\alpha &     f_{v_{1}}    &      \alpha      &       0          \\
    g_{u_{1}}    & g_{v_{1}}-\beta  &         0        &     \beta        \\
     \alpha      &         0        & f_{u_{2}}-\alpha &    f_{v_{2}}     \\
        0        &       \beta      &     g_{u_{2}}    & g_{v_{2}}-\beta
\end{bmatrix}_{U=U^{*}}.
\end{align*}
Let, $\mathbf{\xi}=e^{\lambda t+i(lx+my)}\mathbf{\Phi}$ be the solution of (\ref{4}), then substituting in (\ref{4}) and after simplification, we get
\begin{equation}{\label{5}}
\lambda \mathbf{\Phi} =(\mathbf{F}-k^{2}\mathbf{D})\mathbf{\Phi},~\mbox{where}~k^{2}=l^{2}+m^{2}.
\end{equation}
We are looking for non-trivial $\mathbf{\Phi}$, hence $\lambda$ has to satisfy the following characteristic equation
\begin{equation}{\label{6}}
\mathfrak{D} (k^{2},\lambda)=\mbox{det}(\mathbf{F}-k^{2}\mathbf{D}-\lambda \mathbf{I_{4}})=0.
\end{equation}
Choosing the same coupling strengths i.e., $\alpha=\beta$, we get from (\ref{6})
\begin{equation}{\label{7}}
\begin{aligned}
\mathfrak{D} (k^{2},\lambda) = ((f_{u_{1}}- & \alpha-k^{2}D_{u_{1}}-\lambda) (f_{u_{2}}-  \alpha-k^{2}D_{u_{2}}-\lambda) +f_{v_{1}}g_{u_{2}}-\alpha^{2})\cdot \\
&  ((g_{v_{1}}-  \alpha-k^{2}\sigma D_{v_{1}}-\lambda) (g_{v_{2}}- \alpha-k^{2}\sigma D_{v_{2}}-\lambda) +f_{v_{2}}g_{u_{1}}-\alpha^{2})\\
& -((f_{u_{1}}-  \alpha-k^{2}D_{u_{1}}-\lambda) f_{v_{2}} + (g_{v_{2}}-  \alpha-k^{2}\sigma D_{v_{2}}-\lambda) f_{v_{1}})\cdot \\
& ((f_{u_{2}}-  \alpha-k^{2}D_{u_{2}}-\lambda) g_{u_{1}} + (g_{v_{1}}-  \alpha-k^{2}\sigma D_{v_{1}}-\lambda) g_{u_{2}} )=0.
\end{aligned}
\end{equation}

In this case, sign of $Re(\lambda)$ depends on $D_{u_{1}}, D_{v_{1}}, D_{u_{2}}$ and $D_{v_{2}}$. Here, we choose the same type of conditions on the wave number and the eigenvalue for which we can get $D_{u_{1}}, D_{v_{1}}, D_{u_{2}}$ and $D_{v_{2}}$. For the resonance behaviour, we choose $\lambda$ has a real value $h_{1}$ at $k=k_{1}$, such that $\mbox{max}(Re(\lambda))=h_{1},\forall k$. 
Since at $k=k_{1}$, $\lambda=h_{1}$, hence from (\ref{7}), we get
\begin{equation}{\label{8}}
\mathfrak{D} (k_{1}^{2},h_{1})=0.
\end{equation}
Also at $k=k_{1}$, $Re(\lambda)$ has maximum, so
\begin{equation}{\label{9}}
\lambda'(k_{1})=0.
\end{equation}
 Again, we choose at $k=k_{2}$, $\lambda$ has another real value $h_{2}$, such that the slope of the characteristic curve at $k=k_{2}$ is zero. So, we have 
\begin{equation}{\label{10}}
\mathfrak{D}(k_{2}^{2},h_{2})=0,
\end{equation}
and
\begin{equation}{\label{11}}
\lambda'(k_{2})=0.
\end{equation}

The pair $(h_{1}, k_{1})$ is called the primary `Turing mode' and $(h_{1},k_{1})$ is called as secondary `Turing mode' [see Fig. \ref{fig:TF2}]. After solving the equations (\ref{8})-(\ref{11}) we get the desired values of $D_{u_{1}}, D_{v_{1}}, D_{u_{2}}$ and $D_{v_{2}}$.

\section{Numerical Results}

Based on theoretical results mentioned in the previous section, we have performed extensive numerical simulations. In the simulations, we have considered the equation (\ref{1}) over a square domain of size $200\times 200$ with no-flux boundary condition and a small perturbation around the homogeneous steady state as the initial condition. Throughout the simulations, we plot only $u_{1}$, when the pattern reach at stationary state after neglecting initial transients. We choose $a = 18.0, \sigma = 9.0$ and $b=1.5$ for all the numerical simulations.

\subsection{Strong Coupling}

In this section we consider the case of two Turing modes with the strong coupling $(\alpha=\beta=1.0)$. Now, in order to study the resonant behaviour with strong coupling with light intensity $w$ increased from $0$ to $2.5$. Initially fix $w=0$ and choose the primary Turing mode $k_{1}=0.3,h_{1}=0.2$. To maintain the ratio $3:1$, of wavelengths of the two modes, we choose $k_{2}=0.9,h_{2}=-0.2$ as a secondary Turing mode [see Fig. \ref{fig:TF2}]. Using these $k_{1},h_{1},k_{2},$ and $h_{2}$ after solving (\ref{8})-(\ref{11}), we get the diffusion coefficients: $D_{u_{1}}=0.856,D_{v_{1}}=2.634,$ $D_{u_{2}}=9.199$ and $D_{v_{2}}=39.897$.

\begin{figure}[H]
        \begin{subfigure}[p]{0.47\textwidth}
                \centering
                \includegraphics[width=\textwidth]{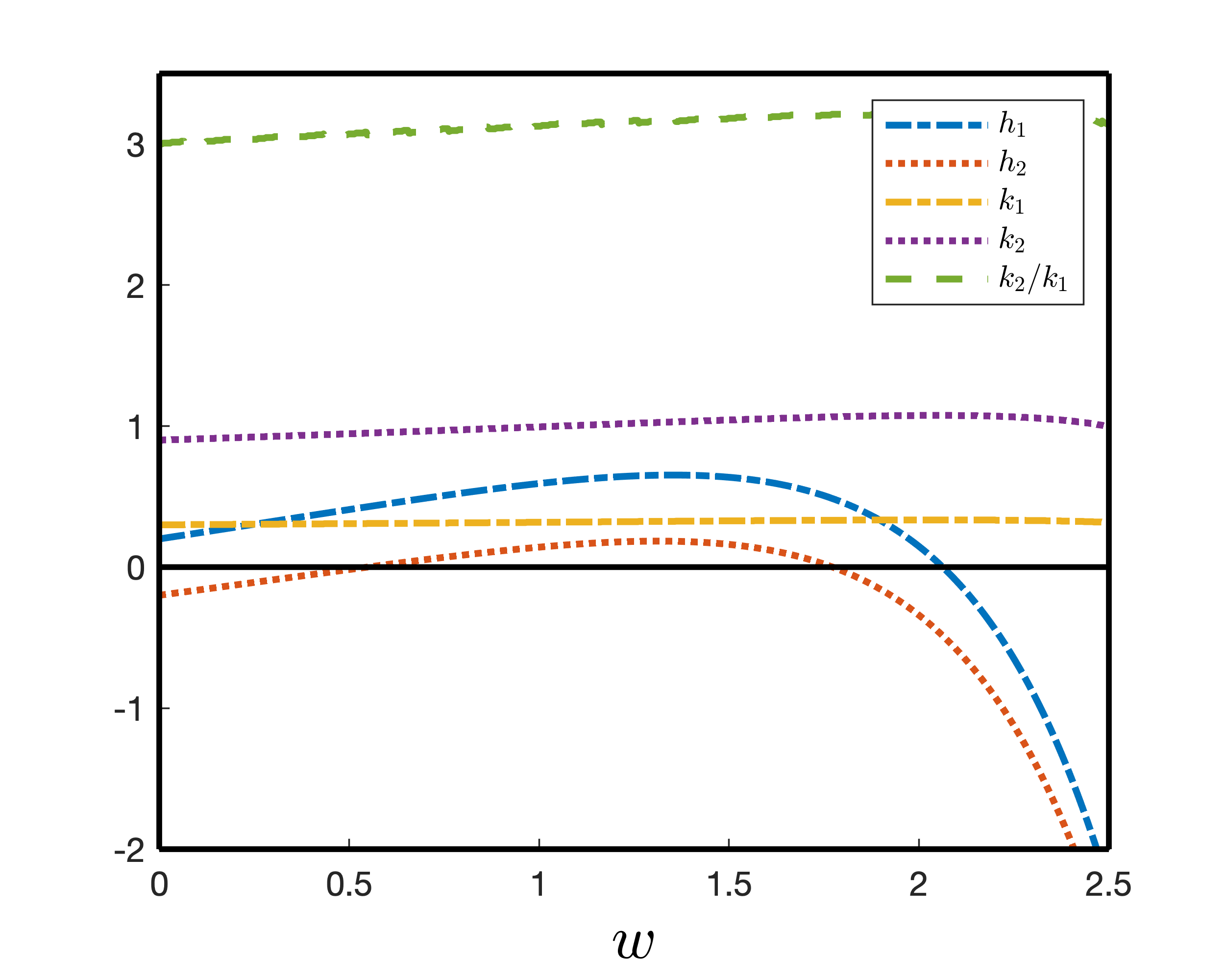}
                \caption{ }
                \label{fig:F1a}
        \end{subfigure}%
        ~~~~~
        \begin{subfigure}[p]{0.47\textwidth}
                \centering
                \includegraphics[width=\textwidth]{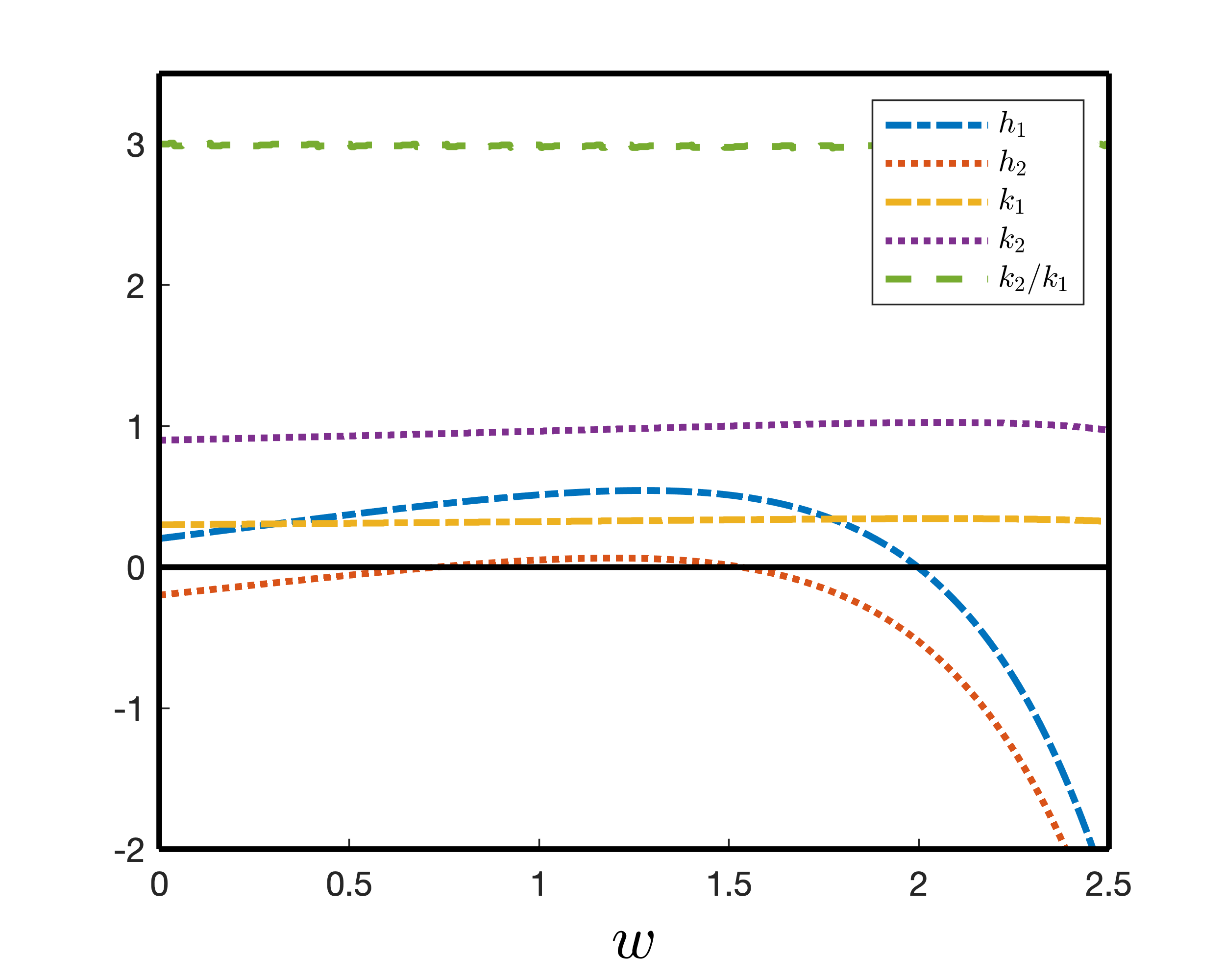}
                \caption{ }
                \label{fig:F1b}
        \end{subfigure}%
        \caption{Plot of the two Turing modes $h_{1},k_{1}$ and $h_{2},k_{2}$ changes for fixed diffusion with increasing $w$: (\subref{fig:F1a}) Strong coupling ($D_{u_{1}}=0.856,D_{v_{1}}=2.634,$ $D_{u_{2}}=9.199$ and $D_{v_{2}}=39.897$)  and (\subref{fig:F1b}) Weak coupling ($D_{u_{1}}=1.404,~D_{v_{1}}=2.088, D_{u_{2}}=10.719$ and $D_{v_{2}}=22.796$).}\label{fig:F1}
\end{figure}

\begin{figure}[H]
                \centering
                \includegraphics[width=\textwidth]{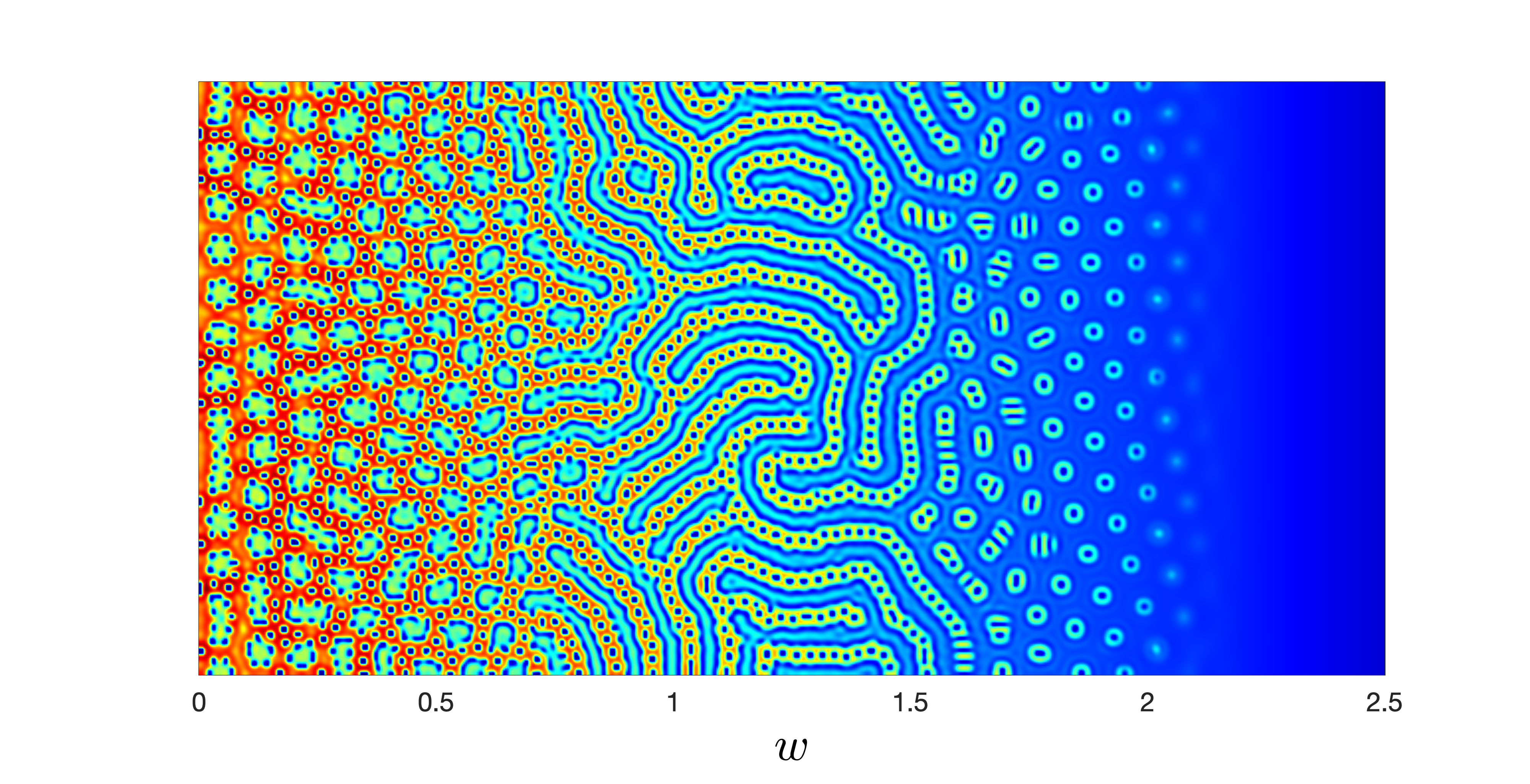}
                \caption{Pattern for the model (\ref{1}) corresponding to the parameter values $\alpha=1, \beta=1, D_{u_{1}}=0.856,D_{v_{1}}=2.634,$ $D_{u_{2}}=9.199$ and $D_{v_{2}}=39.897$.}
                \label{fig:F2}
\end{figure}

For the internal spatial resonance, we require three conditions \cite{Eight} : $(i)$ The ratio of wavelengths of the two modes is close to $\sqrt{3}:1,$ $2:1$ or $3:1$; $(ii)$ the two modes must have the same symmetry; and $(iii)$ the maximum of $Re(\lambda)$ lie within the ranges $0<h_{1}<0.3$ and $-0.5<h_{2}<0$. Now, fixing the diffusion coefficients and solving the system (\ref{1}), the conditions, $(i)$ and $(iii)$ are satisfied, but the  condition $(ii)$ is violated because for $w=0$, we get asymmetric pattern. So, for $w=0$, we get the superposition of the patterns. If we increase $w$ from $0$ to $2.5$, keeping the diffusion coefficient as fixed, then sometimes the hight $h_{1}$ or $h_{2}$ doesn't satisfy the condition $(ii)$ and the ratio of wavelengths of two modes varies in between $3:1$ and $3.2:1$ [see Fig. \ref{fig:F1}(\subref{fig:F1a})]. Also, the internal spatial resonance conditions are violated for $0.24 \leq w \leq 1.9175$. However, only for $1.9175\leq w\leq 2.06$, all the conditions  $(i)$, $(ii)$ and $(iii)$ are satisfied. Again, for $w > 2.06 $ the steady-state of the model (\ref{1}) becomes stable. We find the superposition pattern changes and becomes `Black-Eye' pattern and then dissolved [see Fig. \ref{fig:F2}].

\subsection{Weak Coupling}

For strong coupling, we have seen that superposition pattern may change to the internal spatial resonance due to the increase of $w$. In this section, we are mainly interested to see the effect of $w$ on two Turing modes with the weak coupling $(\alpha =\beta =0.1)$. Choose $w=0$ and assume two Turing modes as $k_{1}=0.3,h_{1}=0.2$ and $k_{2}=0.9,h_{2}=-0.2$. Then the wavelength ratio of the two modes is $3:1$. For this two Turing modes, we have determined: $D_{u_{1}}=1.404,~D_{v_{1}}=2.088, D_{u_{2}}=10.719$ and $D_{v_{2}}=22.796$.

\begin{figure}[H]
        \begin{subfigure}[p]{0.33\textwidth}
                \centering
                \includegraphics[width=\textwidth]{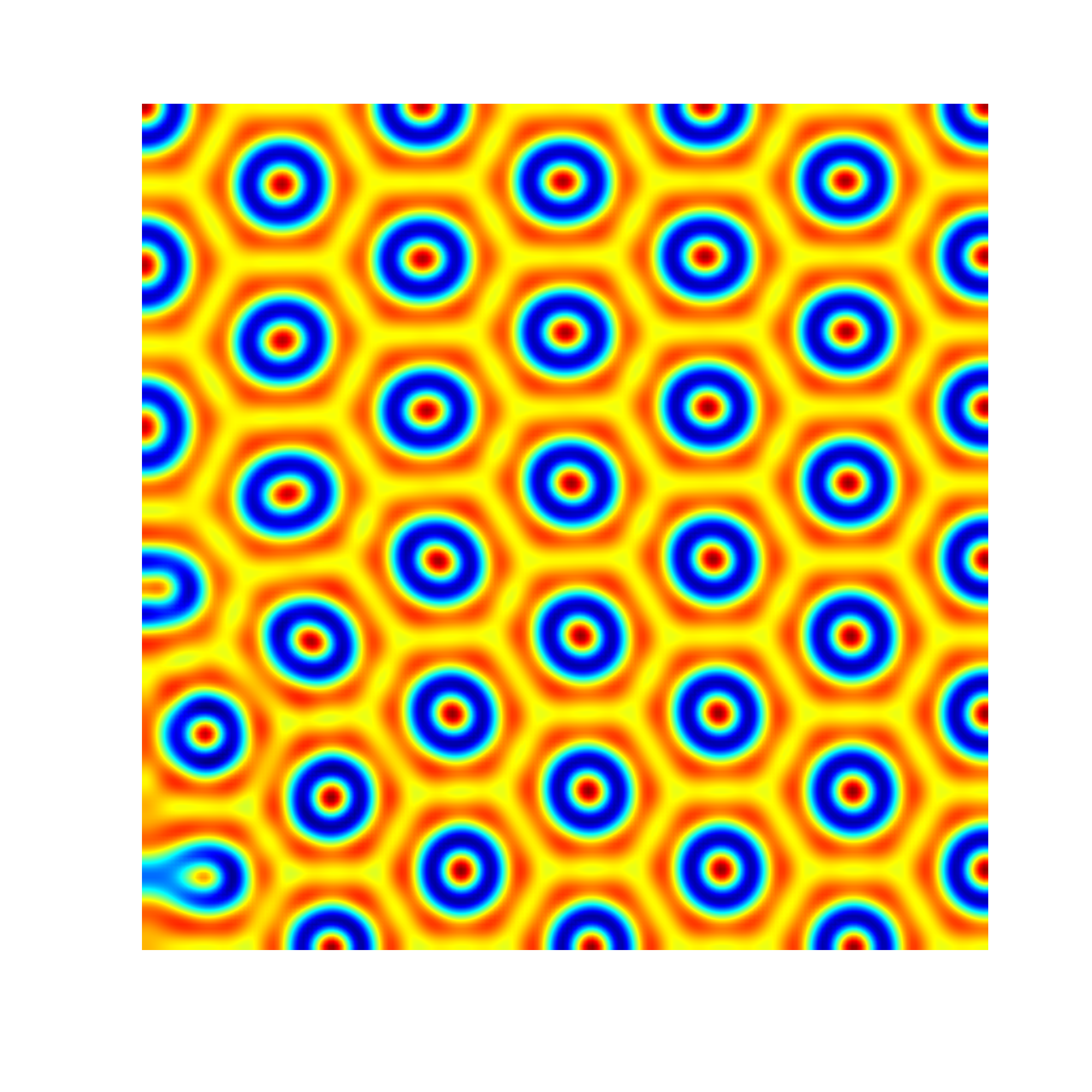}
                \caption{  }
                \label{fig:F3a}
        \end{subfigure}%
        \begin{subfigure}[p]{0.33\textwidth}
                \centering
                \includegraphics[width=\textwidth]{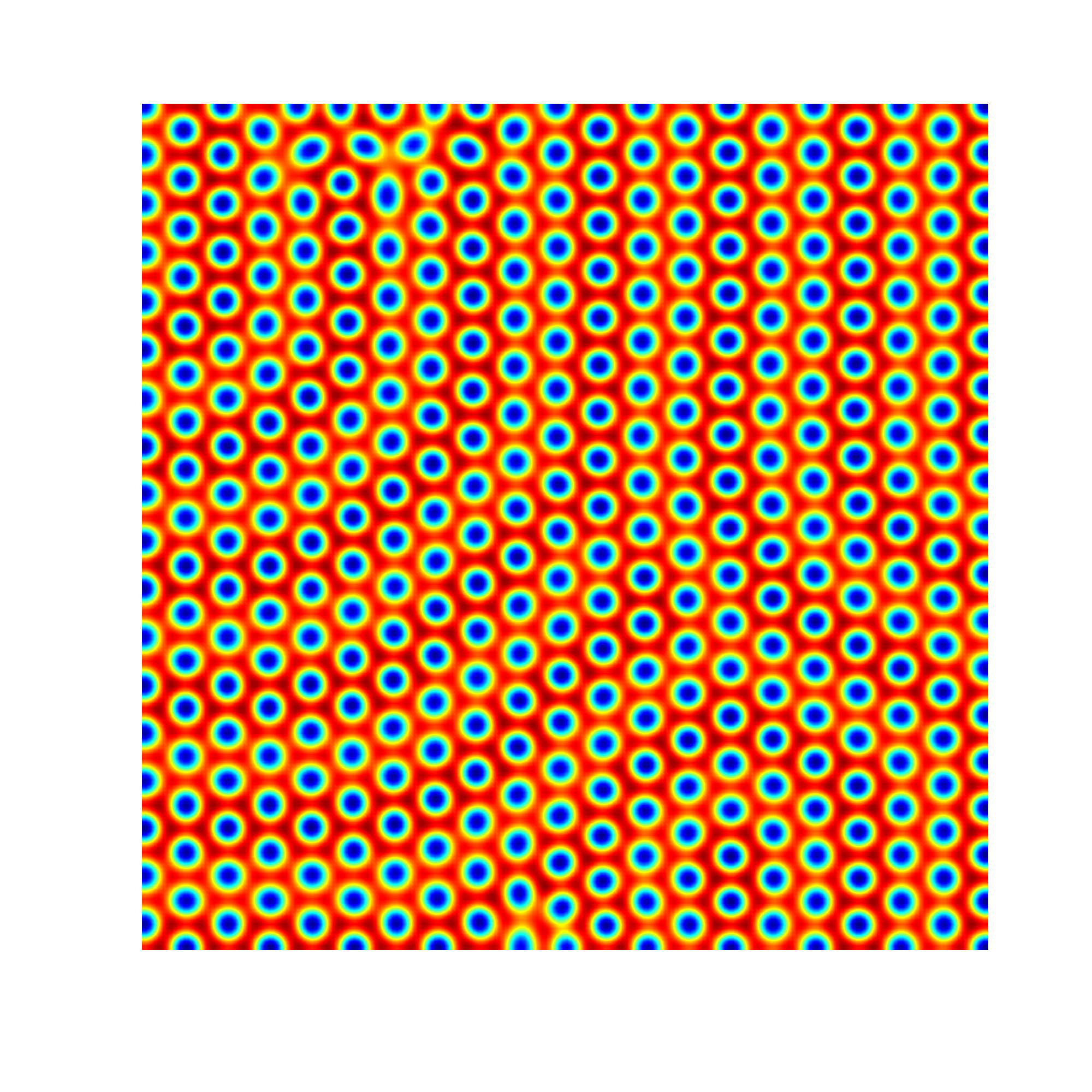}
                \caption{ }
                \label{fig:F3b}
        \end{subfigure}%
        \begin{subfigure}[p]{0.33\textwidth}
                \centering
                \includegraphics[width=\textwidth]{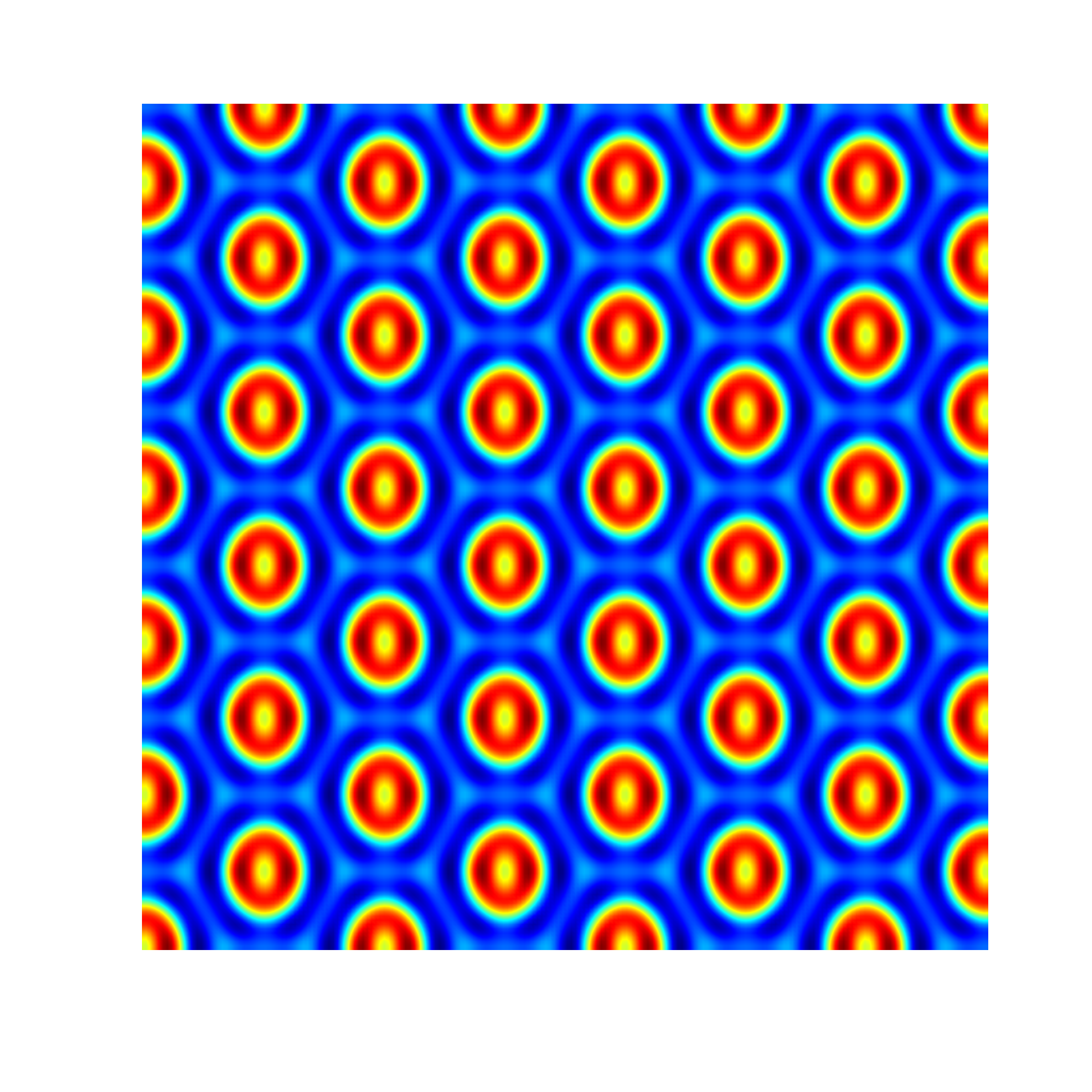}
                \caption{  }
                \label{fig:F3c}
        \end{subfigure}%
\\
        \begin{subfigure}[p]{0.33\textwidth}
                \centering
                \includegraphics[width=\textwidth]{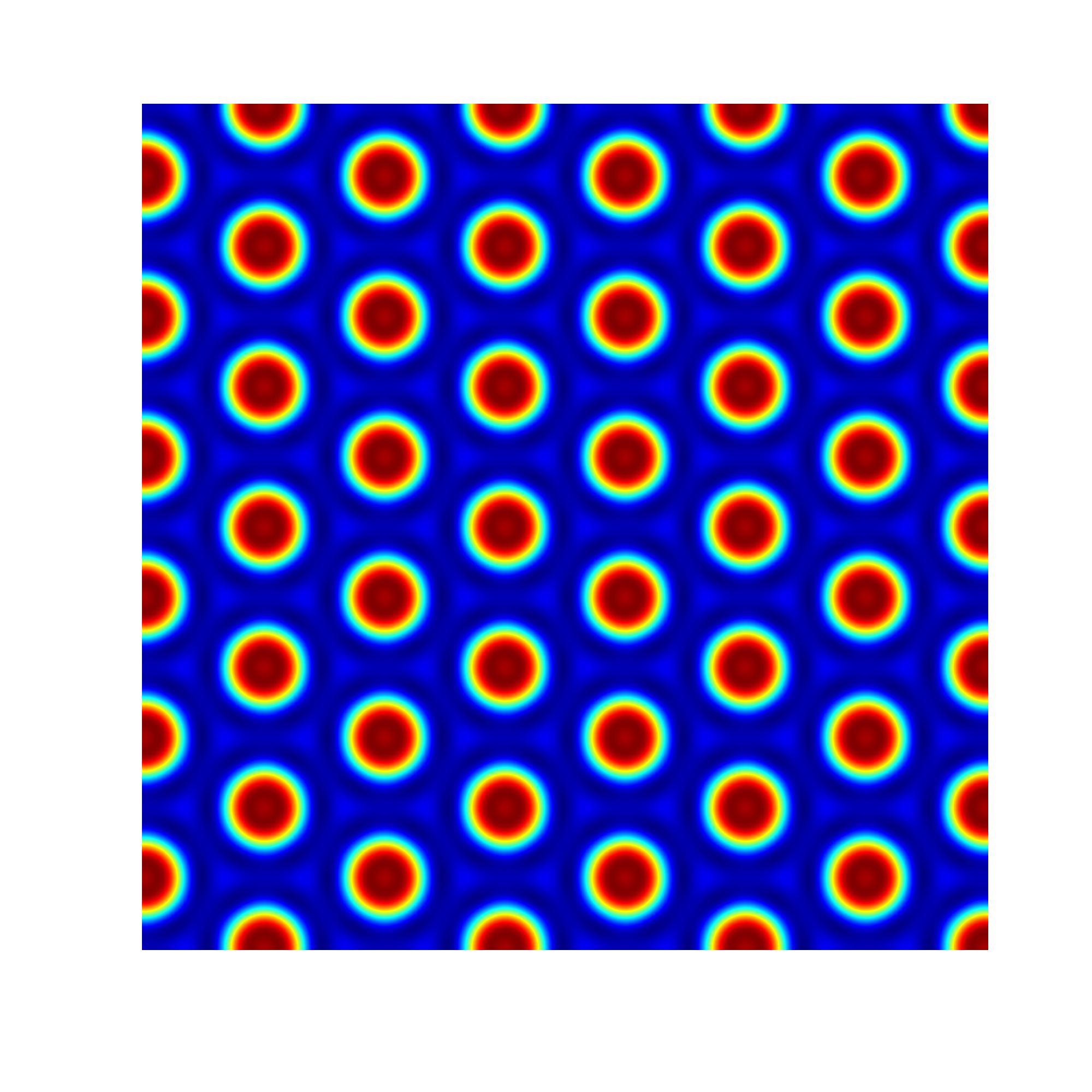}
                \caption{  }
                \label{fig:F3d}
        \end{subfigure}%
        \begin{subfigure}[p]{0.33\textwidth}
                \centering
                \includegraphics[width=\textwidth]{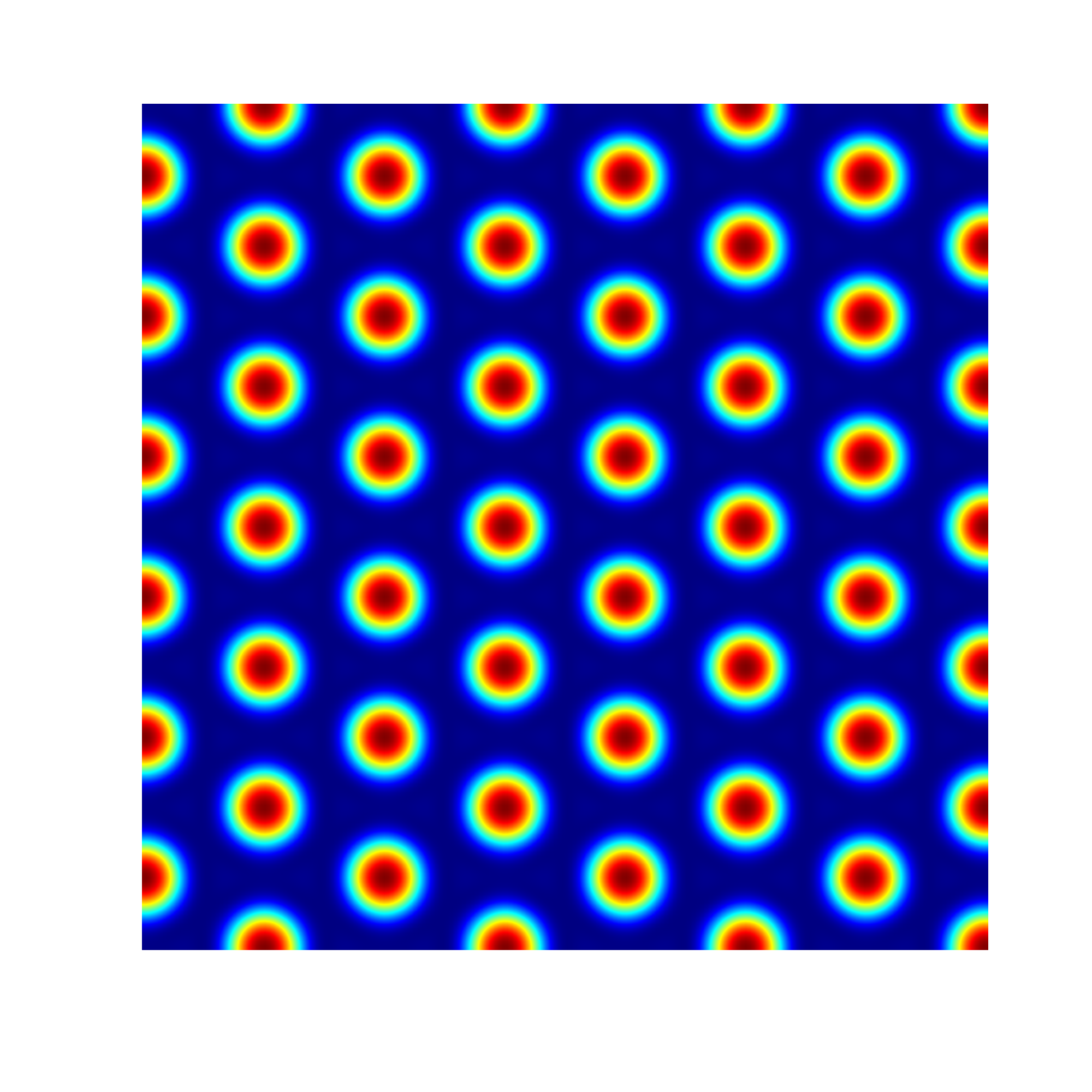}
                \caption{ }
                \label{fig:F3e}
        \end{subfigure}%
        \begin{subfigure}[p]{0.33\textwidth}
                \centering
                \includegraphics[width=\textwidth]{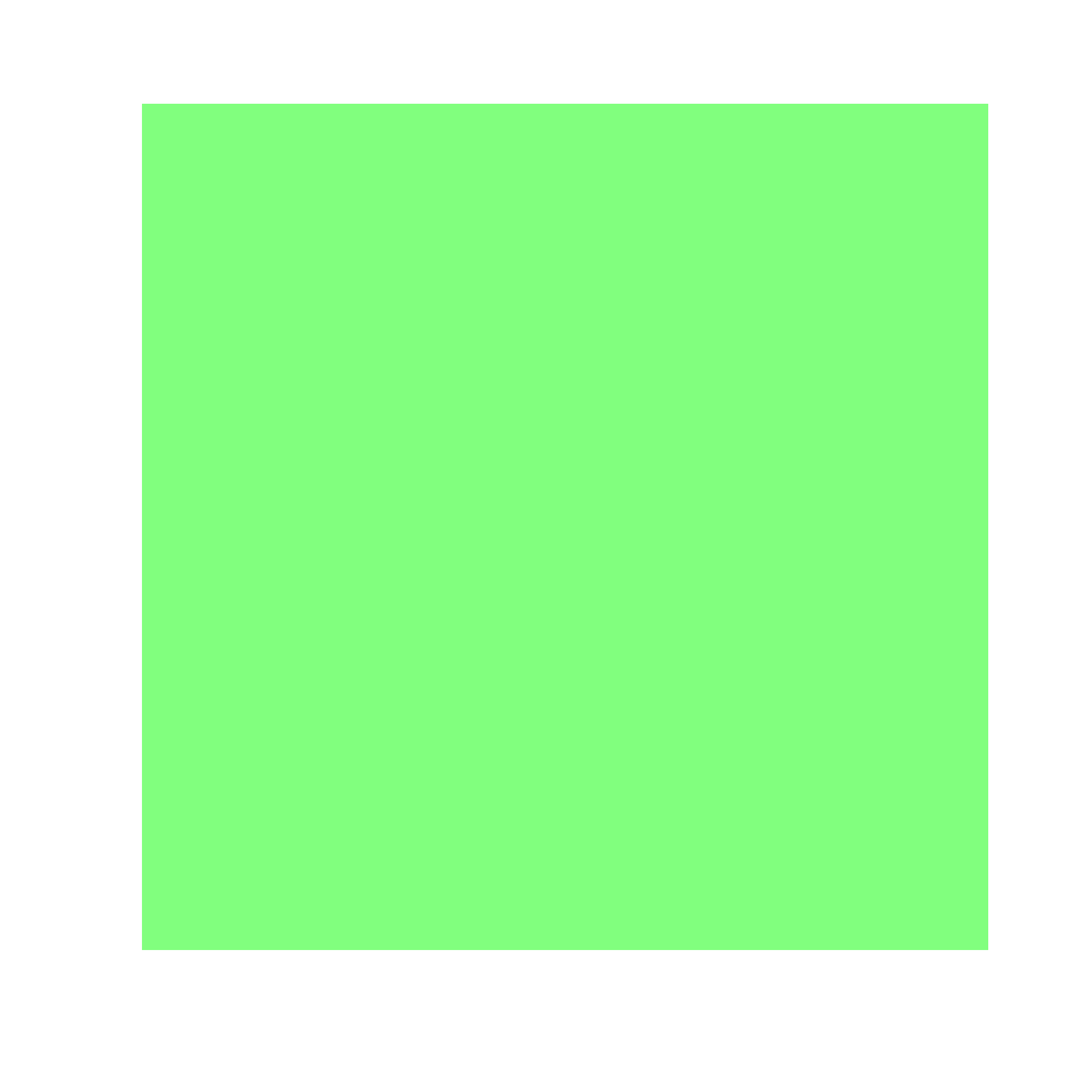}
                \caption{ }
                \label{fig:F3f}
        \end{subfigure}%
        \caption{Patterns for the model (\ref{1}) for the parameter values $\alpha =\beta =0.1$, $D_{u_{1}}=1.404, D_{v_{1}}=2.088, D_{u_{2}}=10.719$ and $D_{v_{2}}=22.796$ with different values of $w$: (\subref{fig:F3a}) $w=0$, (\subref{fig:F3b}) $w=1.0$, (\subref{fig:F3c}) $w=1.7$, (\subref{fig:F3d}) $w=1.87$, (\subref{fig:F3e}) $w=1.99$ and (\subref{fig:F3f}) $w=2.0$.}\label{fig:F3}
\end{figure}

For $w=0$, we find the spatial resonance pattern, composed of black spots in a hexagonal pattern, namely ``honeycomb-type" pattern [see Fig. \ref{fig:F3}(\subref{fig:F3a})]. When $w$ increases from $0$ to $2.5$, the wavelength ratio of the two modes is always close to $3:1$ [see Fig. \ref{fig:F1}(\subref{fig:F1b})].  For $0\leq w\leq 0.285$, all the conditions \cite{Eight}, $(i)$, $(ii)$ and $(iii)$ are satisfied. For $0.285 <w<1.8075$, the condition $(iii)$ is violated. So in this region, we get the superposition patterns. Again, in the region $1.8075\leq w \leq 1.985$ all the internal spatial resonance conditions are satisfied, for $1.985<w\leq 1.995$ the condition $(iii)$ is violated, and after that $(w>1.995)$ the homogeneous steady-steady of the system (\ref{1}) becomes stable.

\begin{figure}[H]
        \begin{subfigure}[p]{0.33\textwidth}
                \centering
                \includegraphics[width=\textwidth]{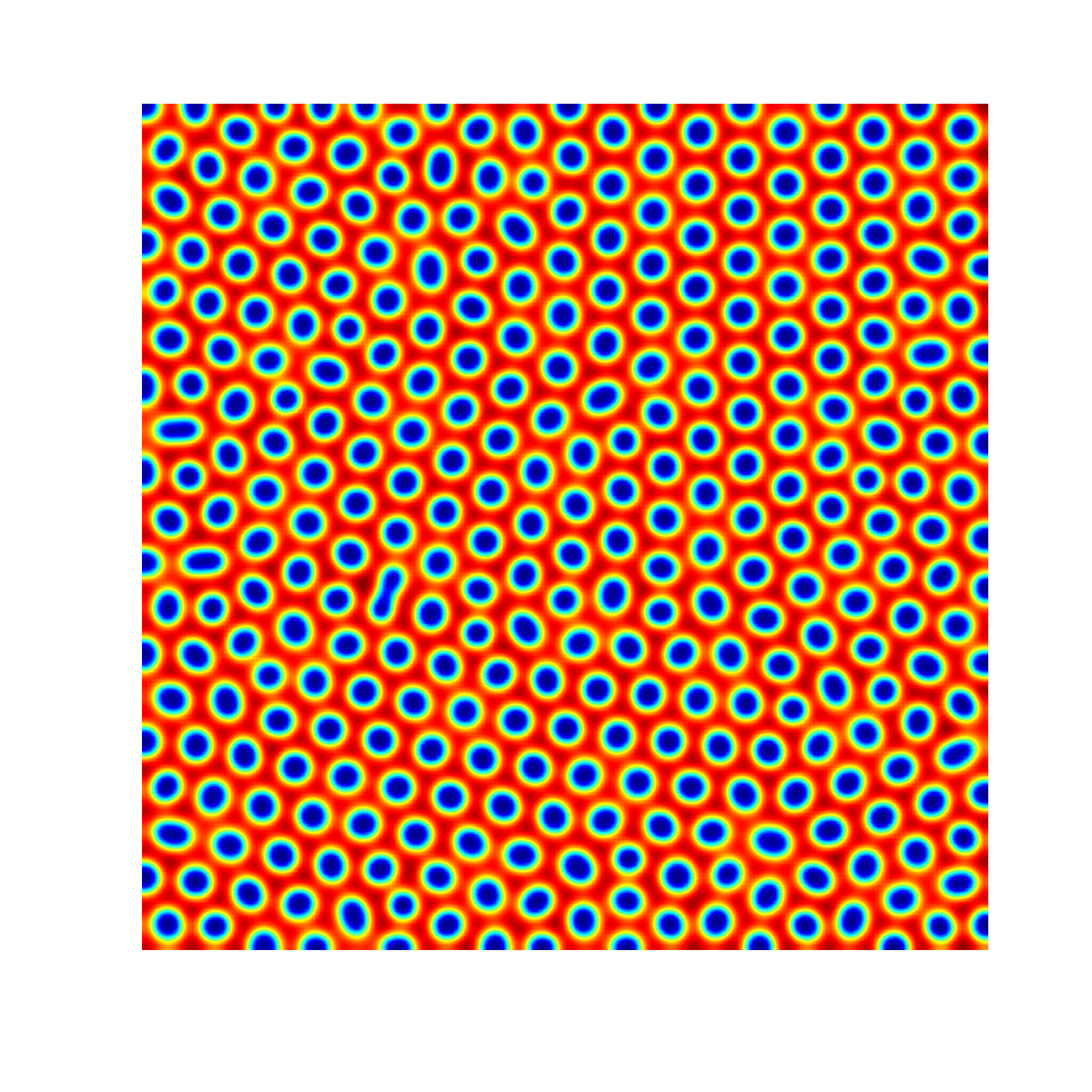}
                \caption{ }
                \label{fig:F4a}
        \end{subfigure}%
        \begin{subfigure}[p]{0.33\textwidth}
                \centering
                \includegraphics[width=\textwidth]{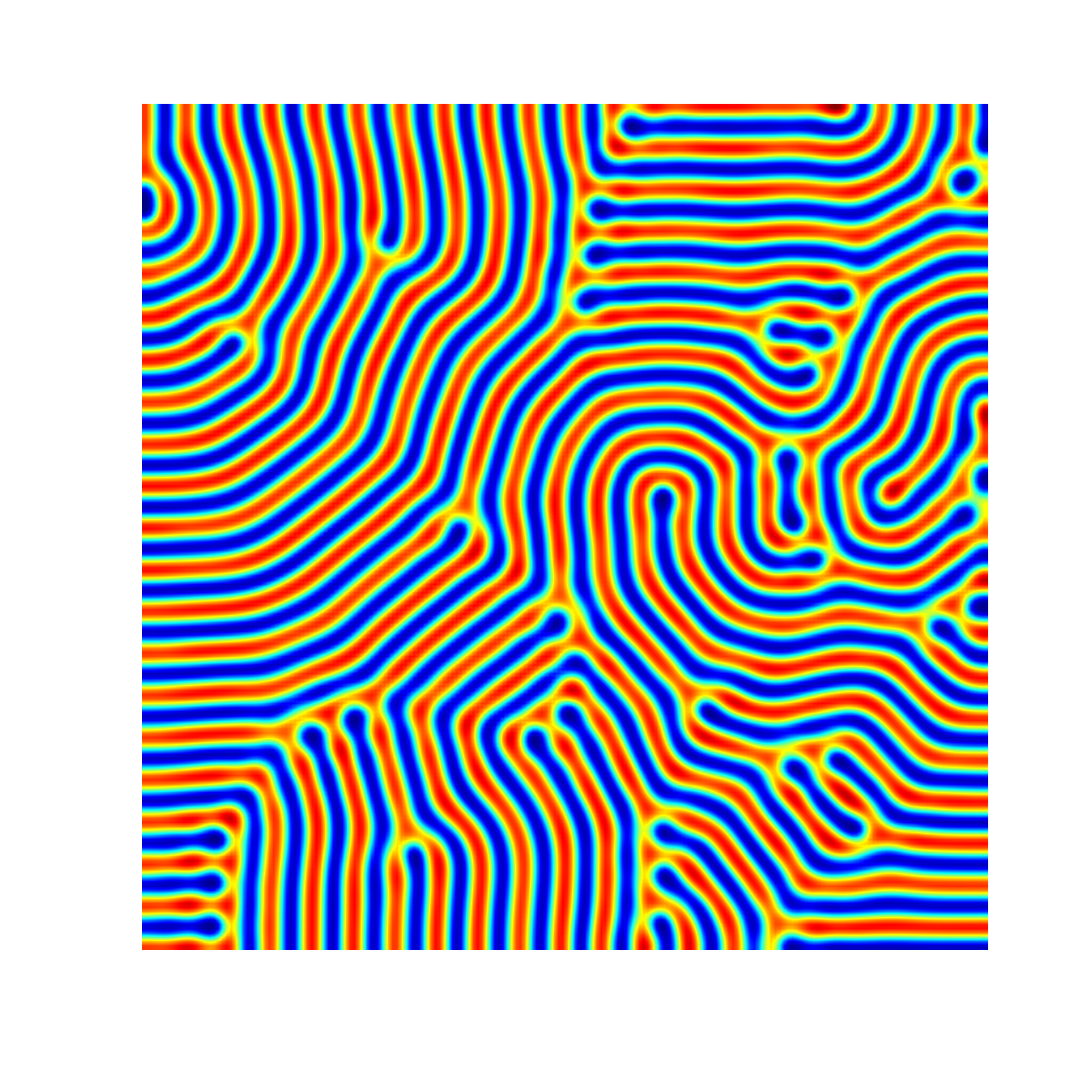}
                \caption{ }
                \label{fig:F4b}
        \end{subfigure}%
        \begin{subfigure}[p]{0.33\textwidth}
                \centering
                \includegraphics[width=\textwidth]{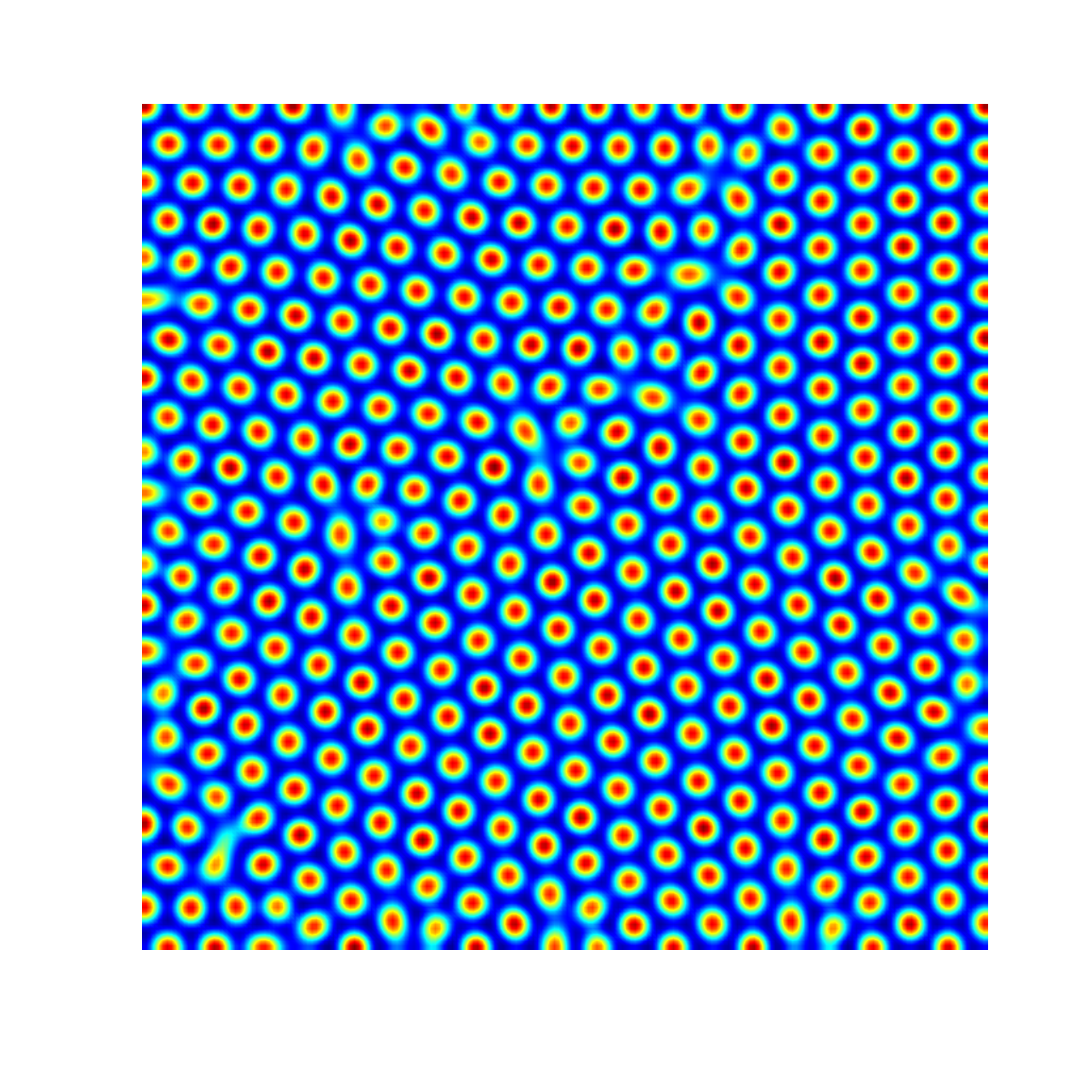}
                \caption{ }
                \label{fig:F4c}
        \end{subfigure}%
        \caption{Patterns for the model (\ref{1}) for the parameter values $\alpha =\beta =0.1$, $D_{u_{1}}=1.270, D_{v_{1}}=2.334, D_{u_{2}}=11.435$ and $D_{v_{2}}=21.098$ with different values of $w$: (\subref{fig:F3a}) $w=0$, (\subref{fig:F4b}) $w=1.0$, (\subref{fig:F4c}) $w=1.8$.}\label{fig:F4}
\end{figure}

Now we choose $w=0$ and assume two Turing modes as $k_{1}=0.3,h_{1}=0.05$ and $k_{2}=0.9,h_{2}=0.05$. Then the wavelength ratio of the two modes is $3:1$. For this chosen Turing modes, we have determined the diffusion coefficients  $D_{u_{1}}=1.270, D_{v_{1}}=2.334, D_{u_{2}}=11.435$ and $D_{v_{2}}=21.098$. If we increase $w$ from $0$ to $2.5$, the wavelength ratio of the two modes is always close to $3:1$. But, for $0\leq w <1.89 $ the condition (iii) is violated. So, in this case we have obtained superposition patterns [See Fig. \ref{fig:F4}]. For $w\geq 1.89$, the homogeneous steady-state of the system (\ref{1}) becomes stable.

\section{Conclusion}

Here, we have studied spatio-temporal pattern formation in CDIMA reaction diffusion system with linear coupling. We have derived the feasibility condition for positive steady-state and Turing instability for both the coupled and uncoupled systems. Here, we have considered both strong and weak couplings and the linear coupling is considered for simplicity. For the coupled system, we have also derived the conditions for critical diffusion coefficients for resonance forcing which is then produce resonance patterns. 

We have verified the spatial resonance patterns via numerical simulations. Different spatial resonance patterns have been observed numerically, depending on the coupling strengths (strong or weak). In the case of strong coupling, superposition patterns appear at zero light intensity. But, an increase in the light intensity, resonance pattern (`Black-Eye') occurs for the reaction-diffusion system, and then homogeneous solution appears. On the other hand, for weak coupling, resonance pattern (`Honey-Comb') exists at the zero light intensity. But increase in the light intensity, it disappears. Sometimes, resonance pattern does not exist and only superposition patterns exist. In this case, it satisfies only $3:1$ resonance forcing but not $h_{1}h_{2}<0$. Finally, along with the resonance forcing $k_{2}:k_{1}$, values of $h_{2}$ and $h_{1}$ is also play an important role in the resulting resonance pattern. Here we have provided the requisite equations by solving which we find the thresholds for diffusion parameter to get resonance patterns.

\end{document}